\documentclass[10pt,twocolumn,twoside]{IEEEtran}

\usepackage[ruled,vlined]{algorithm2e}
\usepackage{graphicx}
\usepackage{subfigure}
\usepackage{url}
\usepackage{cite}
\usepackage{amssymb}
\usepackage{amsmath}
\usepackage{cases}
\usepackage{caption}

\newtheorem{lemma}{\hspace{0.0in}{\bf Lemma}}

\newtheorem{proposition}{\hspace{0.0in}{\bf Proposition}}
\def\done{\hspace*{\fill} \rule{1.8mm}{2.5mm} \\ }

\def\done{\hspace*{\fill} \rule{1.8mm}{2.5mm} \\ }

\begin{document}
\title{Smart Streaming for Online Video Services}
\author{Liang Chen, Yipeng Zhou, Dah Ming Chiu}
\maketitle
\begin{abstract}
Bandwidth consumption is a significant concern for online video service providers. Practical video streaming systems usually use some form of \emph{HTTP streaming} (progressive download) to let users download the video at a faster rate than the video bitrate. Since users may quit before viewing the complete video, however, much of the downloaded video will be ``wasted''. To the extent that users' departure behavior can be predicted, we develop \emph{smart streaming} that can be used to improve user QoE with limited server bandwidth or save bandwidth cost with unlimited server bandwidth. Through measurement, we extract certain user behavior properties for implementing such smart streaming, and demonstrate its advantage using prototype implementation as well as simulations.
\end{abstract}

\section{Introduction} \label{introduction}
A video streaming service lets users download a video file and play it at the same time. At what rate should each user download (stream) the video file? - this is the question we study. Video streaming consumes vast amount of bandwidth in today's network. So this question is of interest to not only the video streaming service provider (also sometimes referred to simply as \emph{content provider}), but anyone concerned about the efficient operation of the Internet as well.

From a content provider's point of view, there are two major concerns - bandwidth cost and user experience (also referred to as Quality of Experience, or QoE). In today's network, a large video streaming service provider typically relies on multiple content delivery networks (CDNs)\cite{agh12unreeling} and spends a large amount of money purchasing the content delivery service. According to our interaction with content providers, the cost can be of the order of 100s million dollars per year. This expense on the bandwidth usage is quite considerable.

The cost of the CDN service is largely determined by the peak bandwidth\footnote{CDN service providers usually charge by either traffic or bandwidth usage. For large volume of video content delivery, they charge by the peak bandwidth usage every month.} the CDN servers use to reach the streaming users. It is a big challenge for content providers to balance the bandwidth consumption and users' QoE. To keep video playback smoothly, the intuitive approach is to deliver surplus video data to users in advance. However, the pre-fetched data at clients will be wasted when users quit viewing videos early (before completing the entire video). How often do users depart and how much bandwidth is wasted? These are the questions we intend to answer in this study.

Since the peak bandwidth cost is roughly determined by the streaming rate per user and the number of served users, a content provider often resorts to either deliver a low bitrate version of the video, or restrict the incoming users at the peak load, to save the bandwidth. Actually, it exchanges the QoE for the bandwidth cost. The video streaming service providers are usually in such a dilemma. Are there better solutions to keep both good QoE performance and reasonable bandwidth consumptions? We target at this issue to design efficient strategies.

Dynamic Adaptive Streaming over HTTP (DASH)\cite{s11mm,s11,lmt12} is an adaptive bitrate streaming technique that enables high quality streaming of media content over the Internet delivered from conventional HTTP web servers\cite{wiki:dash}. It addresses problem of adapting to network congestion, and network bandwidth fluctuation, in addition to adapting to server's bandwidth changes. DASH breaks the content into a sequence of small HTTP-based file segments (or chunks) at multiple resolutions. In the practical implementation, it is an important issue to determine how many segments to order in once request, which is related to the bandwidth usage and QoE performance. Our study intends to support DASH by involving the concern of large amount of bandwidth consumption.

Before DASH is introduced in VoD service industry, the online video content is delivered to users by progressive download \cite{wiki:pd} and other streaming techniques (such as \cite{rtmp} and \cite{mms}). Delivery of a file over HTTP is normally referred to as ``progressive download'' or ``HTTP streaming'' (the trivial difference between them is that progressive download assumes downloading the whole file from the beginning). It is a very simple bulk download of a (part of) video file to end user's computer within the end-to-end bandwidth capacity, and the pre-fetching in local buffer should also help reduce the bandwidth need at peak load. We use the term of ``\emph{streaming}'' to stand for the ``HTTP streaming'' throughout the paper, although it may refer to protocols like RTMP and MMS in history. By ``\emph{smart streaming}'', we mean that a HTTP streaming (including HTTP adaptive bitrate streaming like DASH) strategy gets the bandwidth conservation involved. The purpose of this paper is to study to what extent the wasted downloading can be minimized, and to what extent such \emph{smart streaming} can also be used to minimize peak bandwidth as well as delivering good QoE at the same time.

The rest of the paper is organized as follows. We first measure the bandwidth wastage in a large VoD service system to learn the early departure tendency in Section~\ref{measurement}. Section~\ref{modeling} gives a high-level description of the problem using a simple abstract model, which allows us to describe the space of \emph{smart streaming} solutions depending on the early departure behavior. Based on the measured results and mathematical analysis, we propose a heuristic solution to balance the QoE and bandwidth consumption in Section~\ref{proposal}. Subsequently, we use extensive simulation experiments to evaluate several canonical solutions and compare them in Section~\ref{simulations}. Our results show that smart streaming can significantly out-perform other less sophisticated versions of downloading strategy, and can achieve significantly better QoE than non-progressive strategies. Finally, we briefly describe our experience in implementing our smart streaming strategy in Section~\ref{experiments}, and how well it performs compared to other strategies based on large scale experiments using real implementations. The experimental results valid our analytical and simulation results.


\section{Wastage Measurement} \label{measurement}
\subsection{Data Source}
The previous measurement study \cite{finamore2011youtube} on YouTube had witnessed the significant wastage of bandwidth consumption. We study the bandwidth wastage (i.e., downloaded content not viewed) based on user behavior by collaborating with Tencent Video\cite{qqvideo}, one of the largest video streaming service providers in China. Their VoD service has more than 50 million daily active users, and more than 2 million users online during busy hours (at the time of measurement). Their video content includes movie, TV episodes, music/entertainment video, as well as short clips of news and sports. The video streaming service is delivered over HTTP, served by many servers in multiple CDN providers. With the help of our collaborator, we engineered the client side to report user behavior to a central cloud. When a user finishes or quits viewing a video, the client side makes a record of all the QoE relevant information, including how long the video is watched, how many freezes there are, and for each freeze the start and end time, plus seek and jump events as well. A total of over 550 million sessions are captured. Based on this rich data source, we are able to analyze various user behaviors and system performance. In this part, we will report the result of user \emph{early departure behavior} and the observed \emph{bandwidth wastage}.

\subsection{Methodology}
There are two issues we need to address firstly: (a) How does user experienced performance affect their behavior? (b) How does the content itself affect user departures?

A previous study \cite{Dobrian:2011} reported that the percentage of time spent in buffering (freezes) has a large impact on the user engagement in VoD services. In other words, poor QoE will definitely lead to more early departures. Since we have a very large data set, we can afford to remove all the views that have imperfect QoE. Furthermore, we remove all records that have seek actions during the session, and all records the viewing did not start from the beginning of the video. These removed parts totally account for 19.5\% of raw dataset. As a result, the user behavior we observe can be considered \emph{natural} early departures.

We also observe that there is significant difference in early departure behavior depending on the video's length, and type (e.g. movies and TV programs vs. news and short video clips). Generally, for short videos, users are more likely to complete viewing the whole video. In our study, we focus on the records of movies and long videos (with the length $>$ 30min) in the viewing with perfect QoE, because they consume the most of bandwidth and is the major source of wastage.


\subsection{Early Departure Behavior}\label{departure}
For a view $k$, let $T_k$ be the length of the video viewed, and $L_k$ be the length of the video, we define \emph{viewing ratio} $v_k$ as:
\[
v_k=\frac{T_k}{L_k}
\]
If the viewing ratio is 0, it means the user quits without viewing at all; alternatively a 100\% viewing ratio indicates a user quits after completion. In practice, if the viewing ratio is greater than a threshold, such as 95\%, the user can be considered to have completed viewing the entire video, since the last few percent may correspond to some trailers. Overall, 56.5\% of videos have an average viewing ratio of 50\% or less.

\begin{figure}[htp]
\centering
\includegraphics[width=0.42\textwidth]{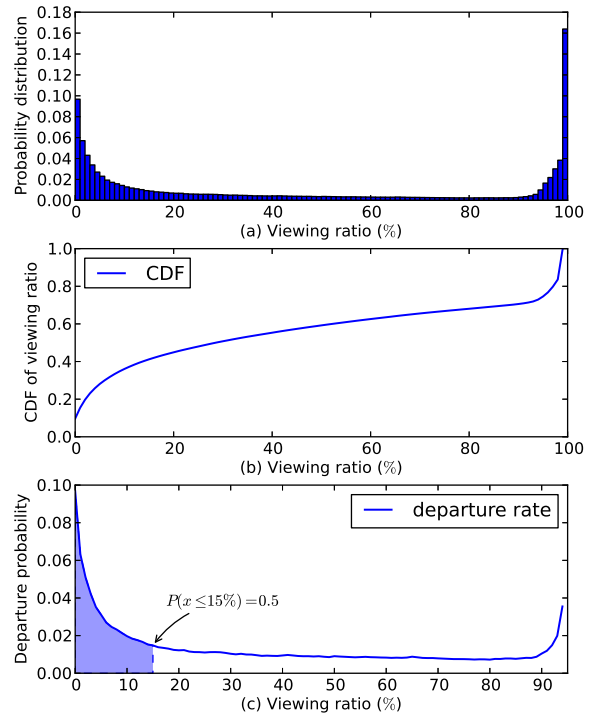}
\caption{Viewing ratio distribution and departure rate distribution.}\label{fig:viewing_ratio}
\end{figure}

We can plot a histogram, with some granularity of viewing ratios, for all movie and long videos. For example, Fig.\ref{fig:viewing_ratio}(a) is such a histogram plotted at granularity of one percent. After plotting this for different hours, days and weeks, we observe that the early departure behavior is quite consistent. The result always comes out more or less the same as shown in Fig.\ref{fig:viewing_ratio}(a).


Our measurements indicate that the departure rate variation over the span of the video is an important factor in designing video streaming strategies. We assume the whole view process is divided into discrete time slots. In any time slot $t$, let $P_t$ be the probability that a user who has viewed $t-1$ time slots of the video will leave (i.e. she will continue with probability $1-P_t$). The third part of Fig.\ref{fig:viewing_ratio} shows $P_t$ plotted against viewing ratio.

Fig.\ref{fig:viewing_ratio}(b) is simply the cumulative distribution for Fig.\ref{fig:viewing_ratio}(a). This will be used to generate random departure times in our simulation and experimentation in later sections. A random number between zero and one can easily be converted to the corresponding random departure time. Fig.\ref{fig:viewing_ratio}(c) is the departure rate distribution, yet another representation of the same information. Let $Q_t$ be the percentage of all users who depart in time slot $t$, the quantity plotted in Fig.\ref{fig:viewing_ratio}(a), and let $P_t$ be the corresponding departure rate in time slot $t$. The relationship between these two quantities are:
\[
Q_t = P_t \prod_{i=1}^{t-1}\left( 1-P_i \right)
\]


From the departure rate histogram, it is clear that users tend to depart with a higher rate at the beginning. Our guess is that movie viewers first go through a movie \emph{browsing} and selection phase, marked by high departure rate. Then they enter a \emph{viewing} phase, during which departure can be caused by a variety of random reasons, and the departure rate remains quite flat, with a spike at the end representing those viewers completing the whole movie. In Fig.\ref{fig:viewing_ratio}(c), we observe that the viewing of videos can be separated into this two phases. Namely, the top departure rates contributing to 50\% of the total departure rates belong to the \emph{browsing} phase, whereas the rest belong to the \emph{viewing} phase. This demarcation can be heuristically used to design our behavior based smart streaming strategy. The idea is, users in the browsing phase have a higher chance of early departure, so we just need to help them download fast enough for good QoE, but not too much pre-fetched data. For users in the viewing phase, however, there is no clear indication to differentiate them from departure rate point of view.

\subsection{Wastage}
\begin{figure}[htp]
\centering
\includegraphics[width=0.42\textwidth]{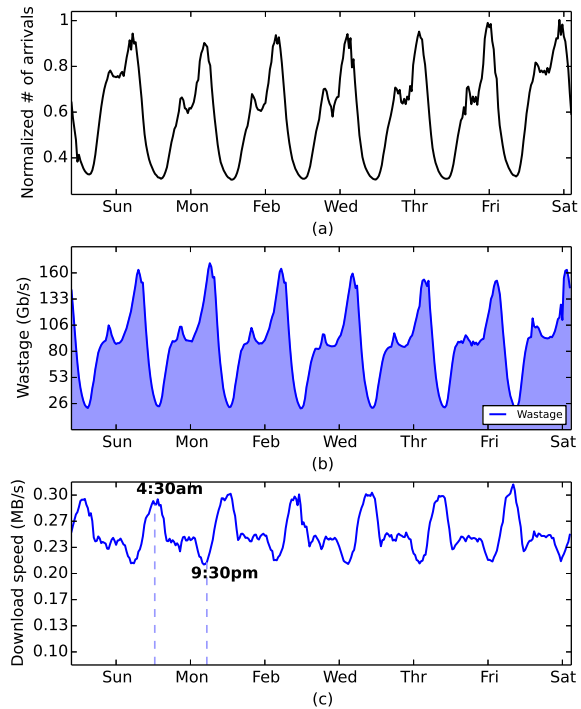}
\caption{Normalized arrivals, the amount of bandwidth wastage per second and the downloading rate measured in one week.}
\label{fig:waste}
\end{figure}

We collected a week's worth of data from our collaborator's VoD system, to examine the bandwidth wastage caused by user's early departure in online video service. The result is shown in Fig.\ref{fig:waste}.

The three plots correspond to (a) the number of arrivals per second (we use normalized values as asked by service provider); (b) the wasted amount of data per second; and (c) the average downloading rate per second. We observed over 20\% of the bandwidth was wasted on average based on the measurement. This result is consistent with previous study on YouTube traffic \cite{finamore2011youtube}. We examine this result because it is helpful for us to appreciate the magnitude of the downloaded content that are not viewed (wasted) due to early departure behavior, hence opportunities for us to do some smart engineering. The three plots together also help illustrate the macroscopic picture of the VoD operation. The average downloading rate (around 2Mb/s) is actually higher than a typical video bitrate rate, indicating the HTTP streaming (progressive download) strategy is adopted. The product of the arrival number and the downloading rate number gives roughly the rate of total server bandwidth consumed, over the week. The difference between downloaded and wasted is the amount of bandwidth used for playback.

\section{Simple Mathematical Analysis}\label{modeling}
\subsection{Assumptions and Notations}\label{Sec:assumption}
Before getting into the engineering details, let us consider the problem at an abstract level, to understand the big picture. For the sake of theoretical analysis, we discuss the resource allocation strategies as there is a central controller, which provides us the idea what an optimal strategy can achieve. In practice, it is implemented and regulated at client side to request the server resource.

Let there be a single server, providing video-on-demand with a fixed uplink bandwidth. All videos have the same length and same playback rate. The video is divided into $L$ segments, each requiring a single time slot to play. In the unit of time slots, $L$ is the length of a video file. The video bitrate is $1$; the server bandwidth is $C$, in the unit of video bitrate.

User requests for a video arrive as a random process. For the sake of simplicity, we assume it is Poisson with rate $\lambda$. All requests for video start from the beginning of the file and proceeds sequentially. Each user has a buffer capable of storing pre-fetched video data. Any surplus (when downloading rate exceeds playback rate) is buffered for subsequent use. During playback, whenever content is missing, it is skipped\footnote{We understand that in practice, most video players would freeze and continue play after sufficient content arrives, which complicates the analysis. In our experimental evaluation later on, we will use a prototype system that freezes instead of skips for missing content.}. With this assumption, the number of users in the system at any time, though random, is determined by a simple stationary process, and will not keep increasing with load (arrival rate). Let $n$ be the random variable denoting the user population, and $N$ denote the expected value. The value of $n$ and $N$ is determined by not only the arrival rate $\lambda$ but also the departure process, we call \emph{user behavior}, described below.


\subsection{User Behavior} \label{userbehavior}
There are three scenarios for user departures:
\begin{itemize}
\item No Early Departure: No user will abort video streaming until the whole video is played. This behavior implies that the viewer population is Poisson distributed with expected value $N=\lambda L$. System load is $\rho = \frac{\lambda L}{C}$.
\item Early Departure with Constant Rate: Each user has the same probability to abort her video streaming session, irrespective how much content has been played. Let $v$ be the fraction of video the user has already viewed. Then, $f(v)$ denotes the probability for that user to abort. In this scenario, $f(v)\equiv \mu$, a constant. This behavior implies that the viewing time is exponentially distributed with average $T = \frac{1}{\mu}$. In this case, the distribution of viewer population is still Poisson, but with expected value is $N=\lambda T$. System load is $\rho = \frac{\lambda}{\mu C}$.
\item Early Departure with Varying Rate: Users abort their video streaming session with different probabilities. At any given time, the departure rate is given by $f(v)$, depending on the fraction of the video already played, the viewing ratio $v$.
\end{itemize}

\subsection{Server Strategies}
Given a particular mix of user video sessions, the server has different options in allocating its bandwidth in serving the user requests. We use the term \emph{streaming} to refer to HTTP streaming, which delivers the video content in the end-to-end throughput capacity. Any surplus will be stored in users' buffers for future playback. Although we use the term \emph{server strategy}, it is understood that in actual implementations, it requires the user end to continue to request for content even when buffered content is already enough to sustain continuous playback, and it requires the users to have ample storage.

By \emph{smart} streaming, we mean rate allocation strategies that progressively use server bandwidth and minimize wasted content simultaneously. The smart streaming strategy is not unique. Among smart streaming strategies, some might be better at minimizing overall skip probability than others. Optimality depends further on modeling of additional buffering can help reduce probability of skipping. The focus of this paper is on \emph{smart streaming}; we will only briefly discuss optimality in our analysis in this section.


Before we get into the analysis, we describe four strategies; together they help characterize the space of different solutions. Note, for each strategy we expect there is some initial buffering before video play starts, and the small buffer (built up by initial buffering) is used to absorb jitters. During initial buffering, video is downloaded as fast as possible.
\begin{itemize}
\item Simple Rate Control (SC): SC tries to maintain the playback rate and not go beyond it. We include it for the purpose of benchmarking. SC does not incur any waste even if the user departs early.
\item Best Effort Streaming (BE): BE corresponds to the way of progressive download which is commonly implemented in HTTP-based streaming. The user end keeps requesting for video chunks. The server tries to respond with best effort, resulting in equal rate when all other things being equal.
\item Equal Buffer Streaming (EB): BE does not take into account of buffer status. Users with ample reserve are treated the same as those with little reserve. EB tries to equalize the reserve for all users. Given initial buffering, the additional reserve may not in itself improve a particular user's QoE. But EB tends to minimize big losses which happens when a user with a big reserve departs early.
\item Equal Waste-Rate (EW): Waste-rate for a user is defined as the buffer length multiplied by the user's departure rate. EW is the strategy that works progressively and equalizes all users' waste rates at the same time. EW is a generalized form of EB.
\end{itemize}
Any of these strategies can be considered \emph{smart} for certain operating scenarios. We give some simple analysis below to fix ideas.

\subsection{The Case of No Early Departure}
\begin{proposition}
If there is no early departure, all strategies are a form of smart streaming.
\end{proposition}

Since there is no early departure, all content pre-fetched at users will be played eventually, so there is no waste; and waste cannot be further minimized.

\subsection{Early Departure with Constant Rate}



In practice, users do depart early. To analyze the situation, we consider a general discrete time model for users with early departures. Because of the stationary (Poisson) arrival process and the skip when no content for playback (rather than freeze) assumption, the number of users into the system is a pre-determined random process independent of the resource allocation strategy. At time slot $t$, let the expected number of users in the system be denoted by $N_t$, and let the expected total amount of content stored at all user buffers be denoted by $S_t$. We can write down $S_{t+1}$ for time slot $(t+1)$ (under heavy load) as follows:
\begin{eqnarray*}
S_{t+1} & = &  S_{t} + C - N_t(1-\gamma) - W_t,
\end{eqnarray*}
where $C$ is the amount of content downloaded by all users in a time slot; $\gamma$ is the average skip probability, and $N_t(1-\gamma)$ is the amount played during time slot $t$. $W_t$ is the expected amount of content wasted at time slot $t$ due to user aborting (or any content arriving late, past the playback point, so it is of no use).
In steady state, $S_{t+1} = S_{t}$ and $W_t$ and $N_t$ become constants denoted by $W$ and $N$, assuming relatively heavy load, that is $N \approx C$. This means:
\begin{eqnarray}
\label{EQ:WandGamma}
W & = &  C - N + N\gamma,
\end{eqnarray}
This can be stated as the following lemma.

\begin{lemma}
\label{LEM:Wandgamma}
Given early departure is with constant rate, the strategy that achieves lower wastage rate $W$ also achieves lower skipping rate $\gamma$.
\end{lemma}

The proof is evident from EQ.~(\ref{EQ:WandGamma}).

As we noted earlier, waste can be due to early departures. The likelihood of skip in practice is hard to model exactly analytically. A reasonable approach is to assume that the skip probability $\gamma_i$ at user $i$ is a function $g(b_i)$, depending on the buffered amount $b_i$ at that user. When a resource allocation policy resulting in buffering state $b_i, (i=1,\dots,N$), then the average skip probability $\gamma$ would be $\sum_{i=1}^{N} g(b_i)$. Furthermore, from intuition the skip probability should monotonically decrease with increase in buffered content. Thus it is reasonable to assume that $g(b)$, i.e. the skip probability function is a convex function of $b$. This implies:
\begin{lemma}
\label{LEM:OPTEB}
Given early departure is with constant rate,
EB minimizes skip probability.
\end{lemma}
\noindent{\bf Proof:}
Since all users have the same departure rate $f(v)\equiv\mu$, $W=\sum_{i=1}^{N} b_{i}\mu$. For any given $N$ users, we would like to determine the amount of stored content at all users $b_i$, ($i=1,\dots,N$) to minimize expected skip probability, given some constant wastage rate $W$. This can be expressed in terms of the following optimization problem:
\begin{eqnarray}
\label{EQ:opt}
\min &  &  \gamma = \frac{1}{N}\sum_{i=1}^{N} g(b_i),\\
 s.t. & & \sum_{i=1}^{N} b_{i}\mu =W.
\end{eqnarray}
Since $g(b_i)$ is a convex function, we should allocate equal buffer $b_i=B$ to all users to achieve the minimum skip probability.
\done

\begin{proposition}
Given early departure is with constant rate, EB is a Smart Streaming strategy.
\end{proposition}
\noindent{\bf Proof:}
First, EB is a HTTP streaming (progressive download) strategy. Secondly, if there is any strategy $\Pi$ that can achieve smaller average skip probability, then from Lemma~\ref{LEM:Wandgamma}, $W_{\Pi}<W_{EB}$. From Lemma~\ref{LEM:OPTEB}, with given $S_{\Pi}$, we can get an even smaller $\gamma$ by allocating buffer resource equally, which is contrary with the optimality of strategy $\Pi$. Thus, EB must be the smart streaming strategy which has minimum $W$.
\done

\subsection{Early Departure with Varying Rates}
In practice, users are likely to depart with varying rates. As will be shown in our measurement section, user departure rate depends on amount of the video already viewed - the more a video has been viewed already, the lower the departure rate.

At a particular time slot, let $v_i$ denote the fraction of video already viewed by user $i$, and $f(v_i)$ denote the departure rate of user $i$. The wastage rate is $W = \sum_{i=1}^N f(v_i) b_i$.
\begin{proposition}
\label{PROP:OPTNEB}
Given early departure is with varying rate, EW is a Smart Streaming strategy, assuming all wastage is due to early departures, and the function $f(v_i)$ is known.
\end{proposition}
\noindent{\bf Proof:}
From the definition of EW, it is evident that EW is a strategy minimizing the wastage.

For a user $i$, the potential wasted content is $f(v_i)b_i$, i.e. the waste rate for user $i$. The wastage rate is $W=\sum_{i=1}^{N}f(v_i)b_i$. Given limited total bandwidth resource $\sum_{i=1}^{N} b_i$, smart strategy should  minimize the expected wasted content, which is equivalent to  equalize the wasted rate vector for all users. Otherwise, we can reduce the wastage rate by allocating more content to the user with less waste rate vector and less content to the user with larger waste rate vector. Thus, EW is a Smart Streaming strategy given varying departure rate.
\done

If we let average skip probability to be a function $g(b)$ depending on the amount of content buffered locally, then the strategy EW is no longer optimal. The following Proposition gives the optimality condition for the most general case:

%
\begin{proposition}
\label{PROP:OPTNEB}
Given early departure is with varying rate, the condition for a strategy to achieve minimized wasted bandwidth  and also minimized skip probability is \[\frac{g'(b_i)}{g'(b_j)} = \frac{f(v_i)}{f(v_j)},\] at any fixed wastage rate $W$ in steady state.
\end{proposition}
\noindent{\bf Proof:}
For any given $N$ users, and varying departure rates, the optimization problem in EQ.~(\ref{EQ:opt}) can be rewritten as:
\begin{eqnarray*}
\min &  &  \gamma = \frac{1}{N}\sum_{i=1}^{N} g(b_i),\\
 s.t. & & \sum_{i=1}^{N} b_{i}f(v_i) =W.
\end{eqnarray*}
Since $g(b_i)$ is a convex function, using Lagrange Multipliers to minimize skip probability, $b_i$ should satisfy: $\frac{g'(b_i)}{g'(b_j)} = \frac{f(v_i)}{f(v_j)}$. This means the users with larger departure rate should be allocated relatively smaller buffered content to minimize skip probability.
\done

In practice, since we do not know $g(b)$ or $f(v)$ exactly, this analysis result could be a high-level guide for us to design a heuristic smart streaming strategy for online video delivery.

To summarize, we have introduced several intuitive resource allocation strategies, in particular EB and EW, besides the strategies used in practice SC and BE.  We have shown that for constant early departure rate, EB is a smart streaming strategy that minimizes waste and skip probability. For variable departure rate, however, EW is only optimal under some idealized assumptions. For practical implementation, we can use the insights gained from the above analysis to build some heuristics.

\section{Heuristic Smart Streaming} \label{proposal}
\subsection{Streaming Strategies Used in Practice}
We first briefly describe the strategies commonly adopted in practice. Most large-scale VoD streaming services today are HTTP-based. The video files are delivered in segments continuously, fetched by the client side with HTTP requests. This framework allows request redirection (the use of multiple CDN servers for service) and load-balancing. The major differences among these HTTP-based streaming are the segment size and the mechanism of bitrate switching.

According to \cite{rao2011network} and \cite{alcock2011application}, Youtube delivers video content in two phases: an initial buffering phase followed by a steady state phase. During the initial buffering phase, the YouTube server serves the video as fast as possible, with the rate limited only by the end-to-end available bandwidth. After initial buffering, the session enters the steady state phase, where the average downloading rate is maintained at the video playback rate plus a increment. In the steady state phase, the video file is delivered in segments. The inter-segment gap is tuned to control the average downloading rate. The size of segment delivered in YouTube is 64KB, and the video resolution is chosen by viewer manually.

\begin{figure}[htp]
\centering
\includegraphics[width=0.3\textwidth]{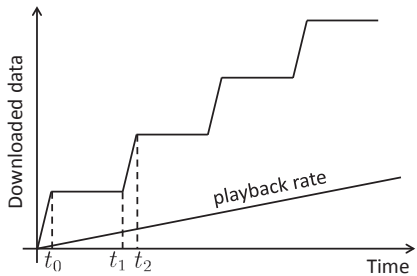}
\caption{Delivery strategy adopted by Tencent Video and Youku, where the 2nd chunk starts downloading at time $t_1$.}\label{fig:youku}
\end{figure}

Also, we measured some popular online video service providers in China, including Tencent Video and Youku. In these VoD systems, long videos are divided into $5\sim7mins$ segments, and the segments are fetched by the client with pauses in between, as shown in Fig.\ref{fig:youku}. Their segment size is much bigger than the segment size in YouTube's case. And many videos they provide tend to be longer than those in YouTube. The switch between different resolutions of video segments is manually controlled by users.

The architecture and service strategy of Netflix was uncovered in the study \cite{agh12unreeling}. Netflix uses the DASH mechanism for video streaming. Each video is encoded at several different quality levels, and is divided into small segments in the length of a few seconds. The client requests several video segments (the range can be adjusted) at each time via HTTP. With each download, it measures the received bandwidth and runs a rate determination algorithm to automatically determine the quality of video segments in the next request.

\subsection{Our Proposed Strategy} \label{BB}
From our measurement results in Section~\ref{measurement}, there are plenty of early departures in VoD streaming sessions. Furthermore, the departure rate varies with the amount viewed, following a function $f(v)$ that is similar in shape for different videos.

From our analysis in Section~\ref{modeling}, however, if the early departure is variable, then the optimal smart streaming strategy requires us to know at least the departure probability function $f(v)$ (in order to implement EW). A more general optimal smart streaming strategy would also require us to know the skip probability function $g(b)$. This would not be a robust solution.

Instead, we propose a heuristic solution based on the salient features of the variable departure rate function $f(v)$ we observed in Fig.~\ref{fig:viewing_ratio}(c). We divide all long videos (including movies) into two phases: (a) a browsing phase with high departure rate, and (b) a viewing phase with roughly constant departure rate. In a practical application, the boundary of two phases depends on the measurements about different types of videos. Based on our observations in Fig.~\ref{fig:viewing_ratio}(c), the top departure rates contributing to 50\% of the total departure rates in the \emph{browsing} phase (viewing ratio less than 15\%) for viewing the long videos.

From our analysis, EB (equal buffer) is already shown to be best for constant departure rate. However, it requires the server to know all user's buffer status and update the resource allocation timely, which is not scalable in the practical implementation. We choose the BE, SC and the mix strategy of them in consideration of their large-scale deployment based on HTTP servers. So the basic idea of our heuristic strategy is that we adopt BE (best effort) for users in the \emph{viewing} phase, and use SC (simple rate control) for users in the \emph{browsing} phase. Since users in the phase of browsing depart more frequently, the server will limit the rate for browsing users to save their bandwidth consumption. And the saved bandwidth can be utilized to support the viewing users to improve QoE performance. We call our heuristic strategy Behavioral-Based Smart Streaming, or simply BB.

Also, we introduce SC+ (plus a delta) to evaluate the performance with SC, BE and BB. For our implementation of SC+, the value of delta is 5\% of video bitrate; in other words, the downloading rate is set to 1.05 times the bitrate. Our focus is the bandwidth wastage and QoE performance that each strategy can achieve.

\section{Simulations} \label{simulations}
Although the abstract analysis in Section~\ref{modeling} helps us to conceptually think about the problem, it is not adequate to convince the practical engineers of our ideas proposed in Section~\ref{proposal}. We shall use both simulation in this section, and experimentation with working prototype in Section~\ref{experiments} to evaluate our smart streaming design. In both cases, we evaluate and compare the streaming strategy designs we introduced in the setting of a VoD server serving a large number of users under heavy load.

\subsection{Simulation Implementation}
For the simulations, we try two different types of user request arrival processes: (1) Poisson arrivals, (2) Trace-driven arrivals. The former gives us a repeatable benchmark, while the latter gives us a glimpse of a more realistic scenario. In both cases, how we set the load level will be discussed later, together with other detailed settings of the simulation. Since the bandwidth wastage is mainly caused by user's early departure behavior, we focus on the the situation of early departure with varying rate as we obtained from our measurement result. We use an extensive set of performance metrics to quantify QoE, based on different statistics related to freezes during playback.

\begin{figure}[htp]
    \centering
    \includegraphics[width=0.4\textwidth]{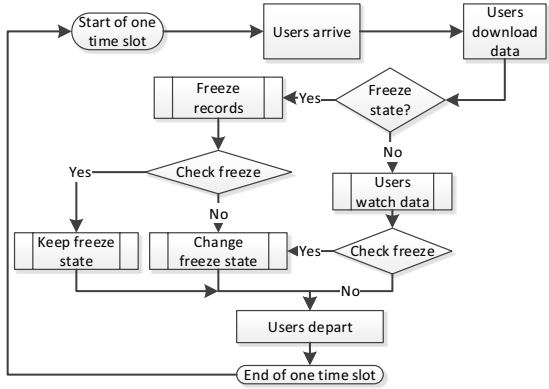}
    \caption{The simulation implementations in unit time slot.}\label{fig:simflow}
\end{figure}

We adopt the activity-oriented paradigm to design our discrete event simulation. Time is divided into time slots, and one time slot represents one second for convenience. In each time slot, our simulation program would go over all the processes sequentially to see if there are activities, as shown in Fig.~\ref{fig:simflow}. For example, in the Poisson arrival case, a Poisson (random) number of users arrivals are generated in each time slot according to workload parameter. For the trace-driven simulation, users arrive according to the real trace data for each time slot. The curve in Fig.~\ref{fig:trace} shows the actual number of arrived users in 86400 time slots of one single day (from one of multiple CDN providers). Given the number of users in the system, the server allocates bandwidth resource following one of the strategies we evaluate, and users get the corresponding amounts of data in that time slot. Users who are not in the freeze state will ``consume'' one time slot amount of the video, while the freezed users will check and update their freeze states according to their buffer status.

For the early departure behavior, we use the measured early departure behavior (as shown in Fig.~\ref{fig:viewing_ratio}(b)) to generate a random departure time for each user. During this process, \emph{playback} is simulated by steadily advancing the playback point (i.e. consuming the video), and checking if the (early) departure point is reached. If either the departure point is reached or the downloading is completed, the concerned user will depart in that time slot.


\subsection{QoE Metrics}

There are no standard metrics for measuring QoE objectively, though there have been studies of various metrics that seem to affect user satisfaction, for example \cite{Dobrian:2011}. We decided to use the following five metrics, to give a fuller coverage. The first three metrics are proposed by our collaborator - they use these metrics as indicators for their system performance. The last two metrics are from \cite{Dobrian:2011}.

\begin{enumerate}
  \item Percentage of users who experienced freeze(s) (PercentUser): This is the proportion of users who experienced at least one freeze during viewing of the video. It indicates the impairment rate of overall QoE performance.
  \item Average number of freezes per user (AvgNFreeze): This is defined as the total number of freezes divided by the total number of users (or sessions), whether they had freezes or not. Those users/sessions still in the system at the simulation end time are not included in the number of total sessions.
  \item Average freeze duration (AvgTFreeze): A freeze starts as soon as local buffer is empty, and ends when a minimum start-up amount (two seconds video data in our simulation) is filled. Average freeze time is defined as total freeze time over the total number of sessions.
  \item Normalized freeze time (FreezeRatio): This is the fraction of the total session time (i.e., playing plus freeze time) spent in freezes.
  \item Rate of freeze events (RateFreeze): FreezeRatio does not capture the frequency of freezes observed by a user, which can be even more annoying than a single freeze of the same amount of time. Thus, this metric is defined as the number of freeze events per minute.
\end{enumerate}

Besides the above metrics, we also include the wasted bandwidth to evaluate the resource usage. We decided not to use the other metrics in \cite{Dobrian:2011} such as rendering quality. Although video quality (resolution or bitrate) is also an important aspect for the QoE in VoD service, our focus in this paper is on the smoothness of playback (as the main QoE performance metric) under different strategies.


\subsection{Simulation Settings}

The key system parameters for all simulation runs are listed in Table \ref{tab:simulation}. We set the video length to be 300 seconds, although it should be much longer in real cases (especially for long videos). This length is enough for the simulation to reach steady state (for the Poisson arrival case), yet not overly long for running the simulations. We also use the \emph{access link rate} to set an upper limit to a single user's downloading rate, to make it more realistic. This means even when a progressive strategy is used, server bandwidth usage is bounded, as assumed in our analysis (EQ.~(\ref{EQ:WandGamma})).

\begin{table}[htp]
    \centering
    \caption{The parameters of simulation setting.}
    \begin{tabular}{l|c}
    \hline
    Video bitrate(CBR):    & 1Mbps\\
    Video length:     & 300s\\
    Access link rate: & 2Mbps\\
    Random seed: & 12345\\
    Server link bandwidth(Poisson): & 1Gbps\\
    \hline
    \end{tabular}\label{tab:simulation}
\end{table}

For the Poisson arrival simulation, we fix the server bandwidth as shown in Table \ref{tab:simulation}. The system's workload is determined by the user arrival rate which we pick. Using the earlier notation, the offered load $\rho$ is given by:
\begin{equation}
\rho=\frac{\lambda L}{C}
\end{equation}
where $C$ is server bandwidth, $L$ is video length, and $\lambda$ is user arrival rate. The value of $\rho$ defines the following load levels:
\[
\rho
\begin{cases}
<1 : & \mbox{light load}\\
\approx1 : & \mbox{heavy load}\\
>1 : & \mbox{overload}\end{cases}
\]
Light load means we have an over-provisioned system, and we expect to have good QoE irrespective of which smart streaming strategy we use. If we are too much into the overload situation, again, smart streaming cannot help. We choose a user arrival rate so that we are in the heavy load case, to see how our strategies can improve system performance. For the no-early-departure case, the choice of arrival rate is easy, straight from the above formula. For the case with early departure, we determine the arrival rate based on the expected viewing ratio, to ensure we are simulating the heavy load scenario.

For the trace-driven simulation, the user arrival rate is given by the trace, and so we pick a \emph{target} server bandwidth instead, to simulate heavy load. In practice, it is also possible that the server bandwidth's limit is not reached. In this case, the higher the peak load bandwidth usage means the higher bandwidth cost the VoD operator has to bear. So we also run an experiment with unlimited server bandwidth, and treat the peak server bandwidth usage as another performance metric for comparison. We will discuss this in more details when we present the simulation results.

\subsection{Poisson Arrival Simulation}
We first take a look at Poisson arrival simulations in the case of early departure with varying rates. For each arrival, we randomly generate the time of departure, using the CDF function of viewing ratio in Fig.~\ref{fig:viewing_ratio}(b). Note, the departure time is not known by the server. In steady state under heavy load, the server's bandwidth is all consumed by the downloading users. The number of users in steady state is determined by the arrival rate and downloading rate. As expected, users always get better QoE under lighter load than that under heavier overload. In fact, at $\rho<0.9$, QoE is perfect. For the overload case $\rho>1$, we see significant degradation in performance.

We compare the different schemes under heavy load ($\rho=0.995$). To make a fair comparison, we are careful to make sure that in all cases we enter the steady state for a similar amount of time. In simulations, users report their session QoE at the time of departure, instead of when completing viewing the whole video. The results are tabulated in Table~\ref{tab:w_simulation}.

\begin{table}[h]
  \centering
  \caption{Comparison of BE, SC and BB algorithms in the case of early departure.}
    \begin{tabular}{|l|c|c|c|c|c|}
    \hline
                  & SC & SC+ & BE & BB \\ \hline\hline
      PercentUser & 86.50\%  & 60.30\%  & 34.74\%  & 15.10\% \\ \hline
      AvgNFreeze  & 1.930    & 0.965    & 0.444    & 0.169 \\ \hline
      AvgTFreeze  & 5.789    & 2.889    & 1.327    & 0.505 \\ \hline
      FreezeRatio & 4.49\%   & 2.25\%   & 1.06\%   & 0.40\% \\ \hline
      RateFreeze  & 0.899    & 0.451    & 0.212    & 0.081 \\ 
    \hline
  \end{tabular}
  \label{tab:w_simulation}
\end{table}

For the SC (simple rate control) algorithm, the rate is controlled to be the same as the bitrate. And SC+ pluses a small delta, as noted before, the 5\% of video bitrate. The bandwidth allocated to each user by SC at a given moment can only be less than that under BE (best effort streaming), and it hardly builds up any reserve. As a result, almost twice the number of users experience freezes, and all the QoE metrics are worse compared to the case for BE.

The implementation of the BB (behavior-based smart streaming) algorithm is as described in Section~\ref{BB}: allocate the rate as same to the video bitrate to those users watching the first 15\% of the video, and apply the rest of the bandwidth to the rest of the users based on best effort streaming strategy. This is assuming the browsing users (the former) are getting a lower rate than the viewing users (the latter); otherwise, rate will be allocated according to BE.

The results in Table~\ref{tab:w_simulation} show that the new algorithm, BB, indeed brings quite a lot of improvement compared to the other algorithms, according to all metrics. This is what we expected. When users depart randomly in the viewing, BB algorithm improves the QoE over others essentially by sensing the buffer/viewing states.


\subsection{Trace-Driven Simulation}
\begin{figure}[htp]
    \centering
    \includegraphics[width=0.45\textwidth]{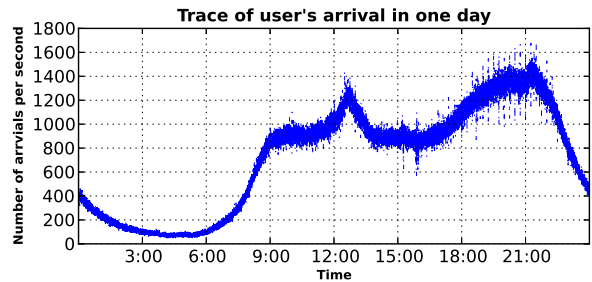}
    \caption{Real trace of user's arrival in one day.}\label{fig:trace}
\end{figure}

We collect the real trace of users' arrival from the VoD service provider, and implement the trace-driven simulation to examine the performance of the four rate allocation strategies. The VoD service provider purchases bandwidth from multiple CDN service providers. Our trace data is based on one of them, which can be taken as a representative sample. The trace of one typical day is shown in Fig. \ref{fig:trace}.

For the case of early departure with varying rates, we implement the trace-driven simulation in two steps. Firstly, we remove the \emph{target} bandwidth setting to examine the peak bandwidth usage of all algorithms. In this case, all users obtain a perfect QoE performance, and the only thing we are comparing is the peak bandwidth usage for the different algorithms. This is shown in the Table~\ref{tab:tracebw}. The peak bandwidth usages of SC and SC+ are only 79.3\% and 81.4\% of that used by BE respectively. The fact the best effort (progressive download) strategies have high peak bandwidth cost is as expected. It is interesting to note that the BB algorithm saves around 9\% of peak load bandwidth compared to other progressive strategies. This is non-trial improvement by concerning both the QoE performance and bandwidth wastage. In industry, the saved percent of bandwidth can help reduce a big chunk of expenses since the purchased bandwidth is always on a large order of magnitude.


\begin{table}[htp]
\centering
    \caption{Trace-driven simulation without the target bandwidth setting.}
    \begin{tabular}{|l|c|c|c|c|}
    \hline
                     & SC & SC+ & BE & BB \\ \hline
      Peak BW(Mb) & 191007 & 196174.7 & 241004 & 220420\\
    \hline
    \end{tabular}
    \label{tab:tracebw}
\end{table}

Secondly, we compare the QoE performance of all algorithms by setting a target bandwidth (181.5Gbps, 95\% peak bandwidth of SC in the previous simulation without limitation). The implementation of the four strategies are the same as that described in the previous simulation, but involving both trace-driven arrivals and bandwidth limit. The results are listed in Table \ref{tab:tracew}. Clearly, the BB strategy gets a significantly better QoE than the other strategies. The differences between SC, SC+ and BE become quite modest in the early departure case. It can be observed BE does not perform as good as SC+ by limiting total bandwidth. The reason can be that too much bandwidth is wasted in BE. Our BB strategy is even better because it utilizes the bandwidth smartly and avoids wastage on those viewers who depart early. Note, it is not meaningful to compare the exact performance levels between the trace-driven and the Poisson cases, since they are not operating under exactly the same heavy load conditions.


\begin{table}[htp]
    \centering
    \caption{Trace-driven simulation for all algorithms in the case with early departure.}
    \begin{tabular}{|l|c|c|c|c|}
      \hline
      & SC & SC+ & BE & BB  \\ \hline
      PercentUser & 15.04\%  & 14.95\%  & 15.02\%  & 6.63\%  \\ \hline
      AvgNFreeze  & 1.664    & 1.597    & 1.643    & 1.116   \\ \hline
      AvgTFreeze  & 5.309    & 5.063    & 5.260    & 3.625   \\ \hline
      FreezeRatio & 3.98\%   & 3.86\%   & 5.37\%   & 3.54\%  \\ \hline
      RateFreeze  & 0.749    & 0.731    & 1.005    & 0.654   \\ \hline
      Wasted BW   & 144.0    & 2465.9   & 18372.6  & 11688.3 \\
    \hline
    \end{tabular}
     \label{tab:tracew}
\end{table}



Also, we added another metric to this table, Wasted Bandwidth, the amount of video data downloaded but not viewed by users over the service duration. This let us further differentiate SC, BE and BB. While BE and BB achieve improved QoE compared to SC, they also \emph{cost} more in terms of wasted bandwidth. Our smart algorithm, BB, improves not only QoE but also wasted bandwidth. In fact, according to our analysis, the two metrics are linked for all smart streaming strategies. That is, BB is able to improve QoE because it reduces the wasted bandwidth.


\subsection{Summary of Simulation Results}
In this section, we have done extensive simulation studies of our proposed behavior based smart streaming algorithm, BB. The ultimate metrics of interests are two folds: QoE and peak bandwidth cost. We can conclude that QoE can be improved over simple rate control (SC or SC+), but at rather high peak bandwidth cost. This cost depends on the design of streaming strategy, the video bitrate chosen by the content provider, and to what extent the access bandwidth of users can limit the peak bandwidth usage. Smart streaming, such as BB, can reduce the peak bandwidth cost, but not as far as what the simple rate control (SC or SC+) can achieve, unless we set a target peak bandwidth for server, as part of the BB strategy. Once we set such a target rate, we expect BB can achieve both improved QoE as well as keep peak bandwidth cost to a level comparable to SC (or SC+).

\section{Experiments} \label{experiments}
In order to further validate our ideas, we implemented a prototype system and ran some experiments. We first describe how we implement the different algorithms in our prototype system, and then the experiments. We had two levels of experimentation. (1) One was a small scale experiment involving a prototype server providing VoD service, and a number of browsers/players running on multiple client machines each accessing the server to play some video. This experiment served to validate our prototype was working correctly in a real environment. The description of how we implemented the HTTP server and the small scope experiment based on HTML5 video is presented in the technical report. (2) In the second experiment, we replaced the real browsers/players with client emulators. The Python emulator allows each physical client machine to emulate dozens of clients making VoD requests, downloading the video and playing it. This enabled us to set up an experiment involving around one thousand emulated clients using the scarce resources we had access to. We describe the large scale experiment in the following subsection.

\subsection{Large Scale Experiment with User Emulators}

\begin{figure}[htp]
        \centering
        \includegraphics[width=0.3\textwidth]{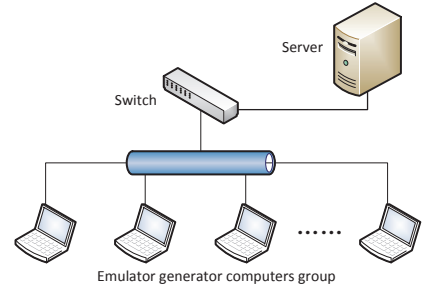}
        \caption{Testbed for large scale experiment.}\label{fig:testbed}
\end{figure}

Our testbed for the prototype experiment consisted of 16 physical computers, one server, one scheduler, and the other 14 running the user emulators. It was a challenge to create the expected arrival pattern visiting the HTTP server in this large-scale experiment. First of all, the time interval of starting two independent user emulators could not be controlled precisely. It was difficult to regulate the 14 machines to generate a workload with Poisson arrival rate. Secondly, it was difficult to manage 14 machines without a central scheduler. So we introduced a scheduler to manage the generation of user emulators, and produce the Poisson number of users per second by each computer in turn according to the expected user arrival rate. Based on this setting, one computer was capable of creating 5$\sim$10 user emulators each second, which met our target well for this experiment.

The emulated users downloaded the video from the server, and reported information back to the server periodically (every 30 seconds). The user emulator did not display the video on the screen. That saved computer resources and allowed thousands of users to be emulated based on the testbed composed of these 14 computers.

\begin{figure}[htp]
        \centering
        \includegraphics[width=0.35\textwidth]{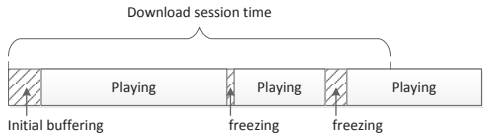}
        \caption{States of user emulator.}\label{fig:states}
\end{figure}

The user emulator is capable of sensing the states as shown in Fig.\ref{fig:states} when they occur with a internal monitoring function. After filling a minimum initial buffer, the user emulator turns into the ``playing'' state and will consume the video data as well as downloading it. If the buffered content becomes less than one second of video, the user emulator enters into the ``freezing'' state and makes a record of it. During the freeze, the emulator continues to download the video data but playback is paused. When the minimal buffer is filled again, the emulated user is back to the ``playing'' state and updates the freeze record. The number of freezes and the duration of freezes are accumulated in one view, and is reported to the server when the emulator departs. The state transitions are shown in Fig.\ref{fig:states}.

When we set up this large scale experiment, we were mindful of possible bottlenecks in the real implementation. For example, we put the entire video file in memory to minimize I/O operations since the server memory was sufficiently large. It was necessary to create almost 70 users for running concurrently in one machine during heavy load. To support its realization, we optimized our emulator code to run as a single process and reduced the resource usage as much as possible. With carefully tuning, we made sure that the CPU, memory and disk loading did not affect the experiment - only the server bandwidth was the bottleneck.



For the case of early departure with varying rate, the user emulator generated the departure time according to the model studied in Fig.\ref{fig:viewing_ratio}(b). With early departures, the Poisson arrival rate in the experiment was set to reach the heavy load situation. To support the implementation, we chose a slightly smaller video bitrate (600Kbps) for this large scale experiment. We adopted the same QoE metrics introduced in Section \ref{simulations} to evaluate the same set of algorithms.

The experimental results are shown in Table \ref{tab:experiment_w}. We run each experiment three times and take the mean of them. Again, we find that BB algorithm achieves better QoE performance than others in the case of early departure. The BB algorithm saves bandwidth and improves QoE, but slightly smaller than that of the simulation, in the experiment with early departure. The relative performance of these strategies are consistent with the simulation results. But the overall experiment performance is poorer than that of the simulations. We recognize that some engineering factors and the design of initial buffer should be considered to optimize the performance in practice.

\begin{table}[tp]
  \centering
  \caption{Performance in experiment for the case with early departure.}
  \begin{tabular}{|l|c|c|c|}
    \hline
      & BE & SC & BB \\ \hline\hline
      PercentUser & 34.3\% & 69.1\% & 21.4\% \\ \hline
      AvgNFreeze  & 1.92   & 7.33   & 0.97   \\ \hline
      AvgTFreeze  & 22.17  & 36.32  & 2.59   \\ \hline
      FreezeRatio & 2.3\%  & 9.1\%  & 0.56\% \\ \hline
      RateFreeze  & 0.31   & 0.78   & 0.21   \\ \hline
      Wasted BW   & 22.5\% & 4.3\%  & 7.1\%  \\
    \hline
  \end{tabular}
  \label{tab:experiment_w}
\end{table}


\section{Related work}\label{relatedwork}
There are quite a few interesting system level and network architectural level studies of VoD streaming services. An interesting work showed the correlation between user engagement and video quality \cite{Dobrian:2011}. They conclude that the time spent in buffering (in this paper we call ``freeze'') has the largest impact on user engagement. The work \cite{liu2012case} proposes to use a global view of client and network conditions to dynamically optimize the video delivery to achieve better QoE for Internet video services. These works help understand Internet-based VoD systems in general. Our work focus on the early departure behavior (not affected by QoE), and the improvement of rate allocation strategy based on real measurement, which is a specific problem not considered in the above works.

The early departure behavior has been considered in some previous works. For example, it was studied as a \emph{preview} activity for users to shop for videos they like \cite{acharya00characterizinguser}, based on data collected on a campus network. Another work \cite{finamore2011youtube} reported statistics of video playback aborts - 60\% of videos watched for no more than 20\% of their duration. The paper \cite{plissonneau2012longitudinal} measures the HTTP streaming traffic from an ISP perspective and find that only half of the videos are fully downloaded. To a large extent, these previous reports agree with our measurement results. Our measurement results are based on data from a large, real-world VoD service provider. The more specific measurement results we got are crucial for implementing our ideas for improving VoD systems. In our previous study \cite{chen2013video}, we also observed users' ubiquitous video browsing behavior in using VoD service.

Other works studied access patterns by users. For example, user arrival patterns are analyzed and modeled in \cite{yu2006understanding}. Some works focus on the transition probability of different user behaviors. In \cite{mongy2007study}, the $K$-means technique is used to retrieve and cluster user behaviors using a Markovian model based on a movie trailer database. Another work \cite{qiu2010analysis} probes the relationship between several types of user behavior and uncovers that the behavior of one individual user in a video streaming session has strong correlation with the user's behaviors in previous streaming sessions. While these works tell us more about user behavior, their results are not particularly helpful in designing our rate allocation algorithms.

There has been many work on adaptive bitrate video streaming. Many earlier works study video streaming over TCP. \cite{wkst08} proposed an analytic performance model to systematically investigate the performance of TCP for both live and stored-media streaming. Paper \cite{kkh10} evaluated and compared three different rate-control algorithms for TCP in terms of the (PSNR) quality of the delivered video and in terms of the timeliness of delivery, which is the first evaluation of TCP-based streaming in an Internet-like setting. More recently, there is a large body of literature on Dynamic Adaptive Streaming over HTTP (DASH), and it has been implemented by industry as well \cite{msstreaming,adobestreaming}. There are many papers discussing how to implement adaptive streaming over HTTP or TCP more effectively and efficiently \cite{s11,ebbghr12}. \cite{s11} provided some insight and background into the DASH specifications as available from 3GPP and in draft version also from MPEG. Papers \cite{xaehej12,xeae13,xaeeh12,lcs10,pmos11,sjk11} analyzed adaptive streaming through stochastic models. The study in \cite{akhshabi2011experimental} measures two major commercial adaptive players and they identify major differences between them. As we explained in the introduction, the problem we study is orthogonal to the work on bitrate adaptation and DASH. They are complementary.

Youtube is one of the largest video provider in the world, and attracted some academic measurement studies. \cite{alcock2011application} investigated the application flow control technique utilized by YouTube. They reveal and describe the basic properties of YouTube application flow control, which is block sending. It also showed that block sending is widely used by YouTube servers. The authors also examined how the block sending algorithm interacts with the flow control provided by TCP. Paper \cite{rlbltd11} studied the network characteristics of the two most popular video streaming services, Netflix and YouTube. Paper \cite{cdl08} crawled the YouTube site for four months, collecting more than 3 million YouTube videos' data. It is reported that Youtube videos have noticeably different statistics compared to traditional streaming videos, ranging from length and access pattern, to their growth trend and active life span.

\section{Conclusion} \label{Conclusion and Future Works}
In this paper, we formulate and study a practical problem for large VoD streaming service providers - how to smartly utilize bandwidth resource to improve streaming QoE and peak load bandwidth requirements. We show that these two goals are highly coupled, and if you can cut down the bandwidth waste you can use the saved bandwidth to improve QoE performance. The key is to understand user early departure behavior. To this end, we collaborate with a large-scale VoD service provider and collected very interesting statistics from their system. The measured statistics let us design a simple and effective rate allocation strategy for our problem at hand. From both simulation and prototyping an experimental system, we demonstrate how our ideas can be implemented in practice and the advantage they can bring to these VoD services.

There are still some facets of interesting work remaining to be pursued. At the analytical end, we think it is possible to create an abstract model to complete analyzing the smart streaming analytically, of course at an abstract level. This would enhance our insights into the fundamental issues of the problem. At the engineering side, our evaluations are based on one single video in short length, which can be extended to support multiple videos in different lengths and different resolutions. Finally, it is also our goal to see our algorithms to be eventually adopted by the industry.

\bibliographystyle{IEEEtran}
\bibliography{mainbib}

\begin{thebibliography}{10}
\providecommand{\url}[1]{#1}
\csname url@samestyle\endcsname
\providecommand{\newblock}{\relax}
\providecommand{\bibinfo}[2]{#2}
\providecommand{\BIBentrySTDinterwordspacing}{\spaceskip=0pt\relax}
\providecommand{\BIBentryALTinterwordstretchfactor}{4}
\providecommand{\BIBentryALTinterwordspacing}{\spaceskip=\fontdimen2\font plus
\BIBentryALTinterwordstretchfactor\fontdimen3\font minus
  \fontdimen4\font\relax}
\providecommand{\BIBforeignlanguage}[2]{{%
\expandafter\ifx\csname l@#1\endcsname\relax
\typeout{** WARNING: IEEEtran.bst: No hyphenation pattern has been}%
\typeout{** loaded for the language `#1'. Using the pattern for}%
\typeout{** the default language instead.}%
\else
\language=\csname l@#1\endcsname
\fi
#2}}
\providecommand{\BIBdecl}{\relax}
\BIBdecl

\bibitem{agh12unreeling}
V.~K. Adhikari, Y.~Guo, F.~Hao, M.~Varvello, V.~Hilt, M.~Steiner, and Z.-L.
  Zhang, ``Unreeling netflix: Understanding and improving multi-cdn movie
  delivery,'' in \emph{INFOCOM, 2012 Proceedings IEEE}.\hskip 1em plus 0.5em
  minus 0.4em\relax IEEE, 2012, pp. 1620--1628.

\bibitem{s11mm}
I.~Sodagar, ``The mpeg-dash standard for multimedia streaming over the
  internet,'' \emph{MultiMedia, IEEE}, vol.~18, no.~4, pp. 62--67, April 2011.

\bibitem{s11}
T.~Stockhammer, ``Dynamic adaptive streaming over http : standards and design
  principles,'' in \emph{Proceedings of the second annual ACM conference on
  Multimedia systems}.\hskip 1em plus 0.5em minus 0.4em\relax ACM, 2011, pp.
  133--144.

\bibitem{lmt12}
S.~Lederer, C.~M{\"u}ller, and C.~Timmerer, ``Dynamic adaptive streaming over
  http dataset,'' in \emph{Proceedings of the 3rd Multimedia Systems
  Conference}.\hskip 1em plus 0.5em minus 0.4em\relax ACM, 2012, pp. 89--94.

\bibitem{wiki:dash}
Wikipedia, ``Dynamic adaptive streaming over http,''
  \url{http://en.wikipedia.org/wiki/Dynamic_Adaptive_Streaming_over_HTTP},
  [Online; accessed 2014-03-09].

\bibitem{wiki:pd}
------, ``Progressive download,''
  \url{http://en.wikipedia.org/wiki/Progressive_download}, [Online; accessed
  2014-03-12].

\bibitem{rtmp}
Adobe, ``Adobe's real time messaging protocol (rtmp),''
  \url{http://www.adobe.com/content/dam/Adobe/en/devnet/rtmp/pdf/rtmp_specification_1.0.pdf}.

\bibitem{mms}
Microsoft, ``Microsoft media server protocol (mms),''
  \url{http://msdn.microsoft.com/en-us/library/cc239490.aspx}.

\bibitem{finamore2011youtube}
A.~Finamore, M.~Mellia, M.~Munaf{\`o}, R.~Torres, and S.~Rao, ``Youtube
  everywhere: impact of device and infrastructure synergies on user
  experience,'' in \emph{Proceedings of the 2011 ACM SIGCOMM conference on
  Internet measurement conference}.\hskip 1em plus 0.5em minus 0.4em\relax ACM,
  2011, pp. 345--360.

\bibitem{qqvideo}
\url{http://v.qq.com/}.

\bibitem{Dobrian:2011}
\BIBentryALTinterwordspacing
F.~Dobrian, V.~Sekar, A.~Awan, I.~Stoica, D.~Joseph, A.~Ganjam, J.~Zhan, and
  H.~Zhang, ``Understanding the impact of video quality on user engagement,''
  in \emph{Proceedings of the ACM SIGCOMM 2011 conference}, ser. SIGCOMM
  '11.\hskip 1em plus 0.5em minus 0.4em\relax New York, NY, USA: ACM, 2011, pp.
  362--373. [Online]. Available:
  \url{http://doi.acm.org/10.1145/2018436.2018478}
\BIBentrySTDinterwordspacing

\bibitem{rao2011network}
A.~Rao, A.~Legout, Y.~Lim, D.~Towsley, C.~Barakat, and W.~Dabbous, ``Network
  characteristics of video streaming traffic,'' in \emph{Proceedings of the
  Seventh COnference on emerging Networking EXperiments and
  Technologies}.\hskip 1em plus 0.5em minus 0.4em\relax ACM, 2011, p.~25.

\bibitem{alcock2011application}
S.~Alcock and R.~Nelson, ``Application flow control in youtube video streams,''
  \emph{ACM SIGCOMM Computer Communication Review}, vol.~41, no.~2, pp. 24--30,
  2011.

\bibitem{liu2012case}
X.~Liu, F.~Dobrian, H.~Milner, J.~Jiang, V.~Sekar, I.~Stoica, and H.~Zhang, ``A
  case for a coordinated internet video control plane,'' in \emph{Proceedings
  of the ACM SIGCOMM 2012}.\hskip 1em plus 0.5em minus 0.4em\relax ACM, 2012,
  pp. 359--370.

\bibitem{acharya00characterizinguser}
S.~Acharya and B.~Smith, ``Characterizing user access to videos on the world
  wide web,'' in \emph{In Proceedings of MMCN}, 2000.

\bibitem{plissonneau2012longitudinal}
L.~Plissonneau and E.~Biersack, ``A longitudinal view of http video streaming
  performance,'' in \emph{Proceedings of the 3rd Multimedia Systems
  Conference}.\hskip 1em plus 0.5em minus 0.4em\relax ACM, 2012, pp. 203--214.

\bibitem{chen2013video}
L.~Chen, Y.~Zhou, and D.~M. Chiu, ``Video browsing -- a study of user behavior
  in online vod services,'' in \emph{The 22nd International Conference on
  Computer Communications and Networks (ICCCN)}.\hskip 1em plus 0.5em minus
  0.4em\relax IEEE, 2013, pp. 1--7.

\bibitem{yu2006understanding}
H.~Yu, D.~Zheng, B.~Zhao, and W.~Zheng, ``Understanding user behavior in
  large-scale video-on-demand systems,'' \emph{ACM SIGOPS Operating Systems
  Review}, vol.~40, no.~4, pp. 333--344, 2006.

\bibitem{mongy2007study}
S.~Mongy, ``A study on video viewing behavior: application to movie trailer
  miner,'' \emph{The International Journal of Parallel, Emergent and
  Distributed Systems}, vol.~22, no.~3, pp. 163--172, 2007.

\bibitem{qiu2010analysis}
F.~Qiu and Y.~Cui, ``An analysis of user behavior in online video streaming,''
  in \emph{Proceedings of the international workshop on Very-large-scale
  multimedia corpus, mining and retrieval}.\hskip 1em plus 0.5em minus
  0.4em\relax ACM, 2010, pp. 49--54.

\bibitem{wkst08}
B.~Wang, J.~Kurose, P.~Shenoy, and D.~Towsley, ``Multimedia streaming via tcp:
  an analytic performance study,'' in \emph{Proceedings of the 12th annual ACM
  international conference on Multimedia}.\hskip 1em plus 0.5em minus
  0.4em\relax ACM, 2004, pp. 908--915.

\bibitem{kkh10}
R.~Kuschnig, I.~Kofler, and H.~Hellwagner, ``An evaluation of tcp-based
  rate-control algorithms for adaptive internet streaming of h. 264/svc,'' in
  \emph{Proceedings of the first annual ACM SIGMM conference on Multimedia
  systems}.\hskip 1em plus 0.5em minus 0.4em\relax ACM, 2010, pp. 157--168.

\bibitem{msstreaming}
\BIBentryALTinterwordspacing
Microsoft, ``Microsoft smooth streaming.'' [Online]. Available:
  \url{http://www.iis.net/download/smoothstreaming}
\BIBentrySTDinterwordspacing

\bibitem{adobestreaming}
\BIBentryALTinterwordspacing
Adobe, ``Adobe http dynamic streaming.'' [Online]. Available:
  \url{http://help.adobe.com/en_US/HTTPStreaming/1.0/Using/index.html}
\BIBentrySTDinterwordspacing

\bibitem{ebbghr12}
J.~Esteban, S.~Benno, A.~Beck, Y.~Guo, V.~Hilt, and I.~Rimac, ``Interactions
  between http adaptive streaming and tcp,'' in \emph{Proc. of ACM NOSSDAV},
  2012.

\bibitem{xaehej12}
Y.~Xu, E.~Altman, R.~El-Azouzi, M.~Haddad, S.~Elayoubi, and T.~Jimenez,
  ``Probabilistic analysis of buffer starvation in markovian queues,'' in
  \emph{INFOCOM, 2012 Proceedings IEEE}.\hskip 1em plus 0.5em minus 0.4em\relax
  IEEE, 2012, pp. 1826--1834.

\bibitem{xeae13}
Y.~Xu, S.~Elayoubi, E.~Altman, and R.~Elzouzi, ``Impact of flow-level dynamics
  on qoe of video streaming in wireless networks,'' in \emph{Proc. of IEEE
  Infocom}, 2013.

\bibitem{xaeeh12}
Y.~Xu, E.~Altman, R.~El-Azouzi, S.~E. Elayoubi, and M.~Haddad, ``Qoe analysis
  of media streaming in wireless data networks,'' in \emph{Proc. of IFIP
  Networking}, 2012.

\bibitem{lcs10}
T.~H. Luan, L.~X. Cai, and X.~Shen, ``Impact of network dynamics on user's
  video quality: Analytical framework and qos provision,'' in \emph{IEEE
  Transcations on Multimedia}, 2010.

\bibitem{pmos11}
A.~ParandehGheibi, M.~M{\'e}dard, A.~Ozdaglar, and S.~Shakkottai, ``Avoiding
  interruptions - qoe reliability function for streaming media applications,''
  \emph{Selected Areas in Communications, IEEE Journal on}, vol.~29, no.~5, pp.
  1064--1074, 2011.

\bibitem{sjk11}
T.~Stockhammer, H.~Jenkac, and G.~Kuhn, ``Streaming video over variable
  bit-rate wireless channels,'' \emph{Multimedia, IEEE Transactions on},
  vol.~6, no.~2, pp. 268--277, 2004.

\bibitem{akhshabi2011experimental}
S.~Akhshabi, A.~Begen, and C.~Dovrolis, ``An experimental evaluation of
  rate-adaptation algorithms in adaptive streaming over http,'' \emph{ACM
  MMSys}, vol.~11, pp. 157--168, 2011.

\bibitem{rlbltd11}
A.~Rao, A.~Legout, Y.-s. Lim, D.~Towsley, C.~Barakat, and W.~Dabbous, ``Network
  characteristics of video streaming traffic,'' in \emph{Proceedings of the
  Seventh COnference on emerging Networking EXperiments and
  Technologies}.\hskip 1em plus 0.5em minus 0.4em\relax ACM, 2011, p.~25.

\bibitem{cdl08}
X.~Cheng, C.~Dale, and J.~Liu, ``Statistics and social network of youtube
  videos,'' in \emph{IWQoS 2008. 16th International Workshop on Quality of
  Service}.\hskip 1em plus 0.5em minus 0.4em\relax IEEE, 2008, pp. 229--238.

\end{thebibliography}

\end{document}